\documentclass[sigconf]{acmart}

\usepackage{algorithmic}
\usepackage{graphicx}
\usepackage{textcomp}
\usepackage{xcolor}
\usepackage[disable]{todonotes}
\usepackage{multirow}
\usepackage{listings}
\usepackage{xcolor}
\definecolor{codegreen}{rgb}{0,0.6,0}
\definecolor{codegray}{rgb}{0.5,0.5,0.5}
\definecolor{codepurple}{rgb}{0.58,0,0.82}
\definecolor{backcolour}{rgb}{0.95,0.95,0.92}

\usepackage{enumerate}

\lstdefinestyle{mystyle}{
  backgroundcolor=\color{backcolour}, commentstyle=\color{codegreen},
  keywordstyle=\color{magenta},
  numberstyle=\tiny\color{codegray},
  stringstyle=\color{codepurple},
  basicstyle=\ttfamily\footnotesize,
  breakatwhitespace=false,         
  breaklines=true,                 
  captionpos=b,                    
  keepspaces=true,                 
  numbers=left,                    
  numbersep=5pt,                  
  showspaces=false,                
  showstringspaces=false,
  showtabs=false,                  
  tabsize=2
}

\colorlet{punct}{red!60!black}
\definecolor{background}{HTML}{EEEEEE}
\definecolor{delim}{RGB}{20,105,176}
\colorlet{numb}{magenta!60!black}

\lstdefinelanguage{json}{
    basicstyle=\normalfont\ttfamily,
    numbers=left,
    numberstyle=\scriptsize,
    stepnumber=1,
    numbersep=8pt,
    showstringspaces=false,
    breaklines=true,
    frame=lines,
    backgroundcolor=\color{background},
    literate=
     *{0}{{{\color{numb}0}}}{1}
      {1}{{{\color{numb}1}}}{1}
      {2}{{{\color{numb}2}}}{1}
      {3}{{{\color{numb}3}}}{1}
      {4}{{{\color{numb}4}}}{1}
      {5}{{{\color{numb}5}}}{1}
      {6}{{{\color{numb}6}}}{1}
      {7}{{{\color{numb}7}}}{1}
      {8}{{{\color{numb}8}}}{1}
      {9}{{{\color{numb}9}}}{1}
      {:}{{{\color{punct}{:}}}}{1}
      {,}{{{\color{punct}{,}}}}{1}
      {\{}{{{\color{delim}{\{}}}}{1}
      {\}}{{{\color{delim}{\}}}}}{1}
      {[}{{{\color{delim}{[}}}}{1}
      {]}{{{\color{delim}{]}}}}{1},
}
\AtBeginDocument{%
  \providecommand\BibTeX{{%
    \normalfont B\kern-0.5em{\scshape i\kern-0.25em b}\kern-0.8em\TeX}}}

\setcopyright{acmcopyright}
\copyrightyear{2023}
\acmYear{2023}
\acmDOI{XXXXXXX.XXXXXXX}

\acmConference[Correctness 2023]{Seventh International Workshop on Software Correctness for HPC Applications}{November 12,
  2023}{Denver, CO}
\begin{document}

\title{DataRaceBench V1.4.1 and DataRaceBench-ML V0.1: Benchmark Suites for Data Race Detection}

\author{Le Chen}
\email{lechen@iastate.edu}
\affiliation{%
  \institution{Iowa State University}
  \city{Ames}
  \state{IA}
  \country{USA}
}
\affiliation{%
  \institution{Lawrence Livermore National Laboratory}
  \city{Livermore}
  \state{CA}
  \country{USA}}

\author{Wenhao Wu, Stephen F. Siegel}
\email{wuwenhao@udel.edu, siegel@udel.edu}
\affiliation{%
  \institution{University of Delaware}
  \city{Newark}
  \state{DE}
  \country{USA}
}

\author{Pei-Hung Lin, Chunhua Liao} 
\email{lin32@llnl.gov, liao6@llnl.gov}
\affiliation{%
  \institution{Lawrence Livermore National Laboratory}
  \city{Livermore}
  \state{CA}
  \country{USA}}

%

\renewcommand{\shortauthors}{Chen, et al.}

\begin{abstract}
Data races pose a significant threat in multi-threaded parallel applications due to their negative impact on program correctness. 
DataRaceBench, an open-source benchmark suite, is specifically crafted to assess these data race detection tools in a systematic and measurable manner.
Machine learning techniques have recently demonstrated considerable potential in high-performance computing (HPC) program analysis and optimization. However, these techniques require specialized data formats for training and refinement.
This paper presents the latest update to DataRaceBench, incorporating new data race contributions from Wu et al. \cite{wu2023model}, and introduces a derived dataset named DataRaceBench-ML (DRB-ML) \cite{drbml}. DRB-ML aligns with the emerging trend of machine learning and large language models. Originating from DataRaceBench, this dataset includes detailed labels that denote the presence of a data race and provides comprehensive details of associated variables, such as variable names, line numbers, and the operation (read/write). Unique to DRB-ML, we have also integrated a series of tailored prompt-response pairs specifically designed for LLM fine-tuning.

\end{abstract}






\maketitle
\todo{add Wu et al.}
\section{Introduction} 
Data races pose a significant challenge in the field of concurrent programming, creating significant obstacles to ensuring software correctness, efficiency, and overall performance. These elusive and intricate bugs can incite non-deterministic and unpredictable behavior, often leading to serious consequences. Liao et al. launched DataRaceBench \cite{liao2017dataracebench} to provide a systematic and quantitative evaluation of these tools. This open-source benchmark suite encompasses diverse data race cases, serving as a valuable asset for researchers and developers. DataRaceBench is publicly accessible on its \href{https://github.com/LLNL/dataracebench}{GitHub repo}. The latest version of DRB before this paper is DataRaceBench V1.4.0, which includes 181 microbenchmarks in C/C++.

Recently, machine learning has showcased remarkable successes across various fields, such as code analysis and parallelization tasks. Nevertheless, a notable gap persists in effectively utilizing and evaluating machine learning in conjunction with DataRaceBench. Applying machine learning techniques, specifically supervised learning models necessitates well-processed and thoroughly labeled data. This facilitates the precise identification of patterns and enables accurate predictions. Furthermore, with the growing interest in deploying Large Language Models (LLMs) for code analysis tasks, there is a clear need for a meticulously curated dataset of prompt-response pairs. These pairs are instrumental in guiding the initial learning process while also providing essential feedback to enhance the model's comprehension and predictive accuracy.

In this paper, we introduce DataRaceBench V1.4.1, an upgraded version of the original DataRaceBench, and DataRaceBench-ML v0.1, a new benchmark derived from DataRaceBench, designed to facilitate the use of machine learning in data race detection. We begin by enriching the DRB with twenty new C code files and updating the corresponding metadata. Building on the foundations of DataRaceBench V1.4.1, we have created DataRaceBench-ML V1.0 with the goal of promoting the development and benchmarking of advanced machine learning methods for data race detection, with a particular focus on Large Language Models. Our contributions are as follows:
\begin{enumerate}[-]
    \item We examined open-source codes from relevant data race studies and adapted them to the DataRaceBench format, thereby extending DataRaceBench with twenty new data files.
    \item We processed DataRaceBench to incorporate informative labels suitable for machine learning-based data race detection and analysis.
    \item We generated prompt-response pairs as part of the DRB-ML dataset to facilitate LLM fine-tuning.
\end{enumerate}


\section{DataRaceBench V1.4.1} 
DataRaceBench~\cite{liao2017dataracebench}, first introduced by Liao et al., has continually evolved, most recently leading to version V1.4.0 ~\cite{lin2021high}. This paper presents the updated DataRaceBench V1.4.1. We delve into the modifications of this enhanced version, detailing the new code inclusions and their implications in the pursuit of optimized data race detection.

\subsection{Origins of Newly Added Microbenchmarks}
\label{sec:process}
Wu et al. \cite{wu2023model} proposed a straightforward model-checking technique that validates a program is free from data races. To evaluate the efficacy of the proposed verifier, they employed DataRaceBench in conjunction with several supplemental examples. We selected twenty C-language programs from these supplemental instances, processed them thoroughly, and ultimately incorporated them into DataRaceBench.

\subsection{Data Processing}
Adhering to the design principles of DataRaceBench, we thoroughly processed the newly incorporated microbenchmarks. Our processing steps encompassed the following:
\begin{itemize}
    \item We examined the length of each microbenchmark, ensuring it is as small as possible to represent typical data race detection cases. The twenty new microbenchmarks did not significantly alter the overall size of the DataRaceBench suite. The average size of microbenchmarks in DRB v1.4.1, measured by the string length, increased from 29.4 to 30.9 compared to DRB v1.4.0. 
    \item We confirmed that each program was self-contained and included a main function, thereby ensuring its independence and usability for both static and dynamic analysis tools.
    \item We categorized each program into either the race-yes or race-no set based on its characteristics.
    \item For programs categorized in the race-yes set, we identified and marked the pair of source locations responsible for causing data races.
    \item Finally, we assigned appropriate data race property labels to the collected microbenchmarks following the DRB design. These labels are used to categorize different types of code patterns, such as unresolvable dependencies, missing data sharing clauses, missing synchronizations, and so on.  
\end{itemize}

\subsection{New Microbenchmark Example}
Listing \ref{lst:ex193} shows an example of new microbenchmarks (DRB193) in DRB v1.4.1. Following the processing steps discussed in section \ref{sec:process}, we transform the data race-related code to a DRB microbenchmark.  DRB193 is a positive benchmark with known a data race pair due to different critical section names used.

\begin{lstlisting}[language=c, caption=DRB193-critical-section3-yes.c, label=lst:ex193]
/*
!!!~~~~~~~~~~~~~~~~~~~~~~~~~~~~~~~~~~~~~~~~~~~~~~~~~~~!!!
!!! Copyright (c) 2017-20, Lawrence Livermore National Security, LLC
!!! and DataRaceBench project contributors. See the DataRaceBench/COPYRIGHT file for details.
!!!
!!! SPDX-License-Identifier: (BSD-3-Clause)
!!!~~~~~~~~~~~~~~~~~~~~~~~~~~~~~~~~~~~~~~~~~~~~~~~~~~~!!!
 */

/*
 * This is a program based on a dataset contributed by
 * Wenhao Wu and Stephen F. Siegel @Univ. of Delaware.

 * Race due to different critical section names
 * Data race pair: x@26:7:W vs. x@43:7:W
 */

#include <stdio.h>
int main()
{
  int x = 0, s = 0;
#pragma omp parallel sections shared(x, s) num_threads(2)
  {
#pragma omp section
    {
      x = 1;
#pragma omp critical(A)
      {
        s = 1;
      }
    }
#pragma omp section
    {
      int done = 0;
      while (!done)
      {
#pragma omp critical(B)
        {
          if (s)
            done = 1;
        }
      }
      x = 2;
    }
  }
  printf("%d\n", x);
}

\end{lstlisting}



\section{DataRaceBench-ML V0.1} 
DataRaceBench  has been proven to be a valuable resource for data race study. However, a gap exists between its current offerings and the specific needs of machine learning approaches designed for data race detection. To bridge this gap, we introduce DataRaceBench-ML (or DRB-ML) V0.1 in this section. DRB-ML is a derivative of DRB specifically crafted to support machine learning methodologies in the realm of data race detection.

\subsection{Label Extraction}
The efficacy of machine learning models, particularly supervised learning, hinges substantially on the quality and detail of the labels associated with the training data.  In the context of data race detection, labels convey critical information, such as whether a data race exists within a given code segment and, if so, the precise location and nature of the race. A dataset with diverse labels can cater to many tasks for machine learning models, enhancing their training and ultimate performance.

\begin{figure}[h]
  \centering
  \includegraphics[width=\linewidth]{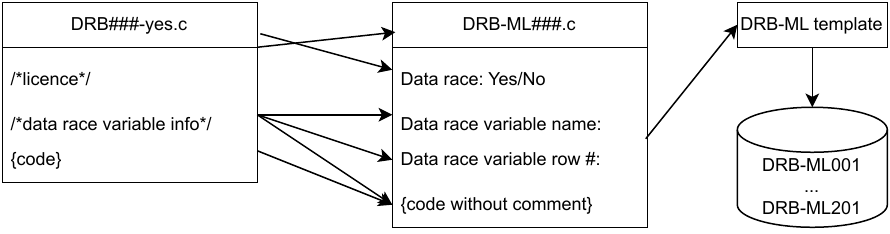}
  \caption{Transformation Process from DRB to DRB-ML Dataset:1. Data Race Status Extraction; 2. Retrieval of Data Race Variables and Lines; 3. DRB Code Comment Trimming; 4. DRB-ML Template Integration.}
  \label{fig:drb2drbml}
\end{figure}

To facilitate the application of machine learning to DataRaceBench, we executed a comprehensive label extraction process. Figure \ref{fig:drb2drbml} showcases the information obtained from DRB to generate labels in DRB-ML. Each data entry in DRB-ML is formatted in JSON and includes the following keys:
\begin{itemize}
    \item "ID": A unique index number starting from 1. It preserves the original ordering in DRB and is stored as an integer.
    \item "name": The original filename of the DRB file. This is directly copied from the DRB filenames and stored as a string.
    \item "DRB\_code": The original code present in DRB microbenchmarks. It is copied verbatim from the DRB benchmarks and stored as a string.
    \item "trimmed\_code": The DRB\_code with all comments removed. We employed a script to process each microbenchmark in DRB and store the comment-free code as a string.
    \item "code\_len": An integer value representing the string length of the trimmed code. This label aids in thresholding data to match the input size constraints of various machine learning models, ensuring compatibility and adaptability.
    \item "data\_race": This is a boolean value representing the data race condition in DRB. The presence of a data race is indicated by 1, and its absence by 0.
    \item "data\_race\_label": This label indicates the type of data race condition (race-yes or race-no) that DRB marks. We used DRB metadata to generate this label and stored it as a string variable.
    \item "var\_pairs": This is a list of pairs of variables associated with a data race. It is empty when "data\_race" is 0. Each item follows the format [VAR0, VAR1], where VAR1 depends on VAR0, and each variable has the following attributes:
    \begin{itemize}
        \item "name": Variable names stored as a string variable.
        \item "line": The line number indicating the variable's location, stored as an integer variable.
        \item "col": The column number showing the variable's location, stored as an integer variable.
        \item "operation": The operation performed on the variable, stored as a string variable. The value is either "w" (representing a write operation) or "r" (representing a read operation). 
    \end{itemize}
\end{itemize}


We automated the extraction process to ensure consistency and scalability. Leveraging a custom-developed script, we systematically traversed through each microbenchmark in DRB, extracting the pertinent information and transforming it into an effective format for machine learning. This process yielded a comprehensive and richly labeled dataset, making it a valuable resource for developing and evaluating machine learning models focused on data race analysis.


\subsection{Prompt-Response Pairs For Supervised Fine Tuning LLMs}
In the context of fine-tuning Large Language Models (LLMs), using prompt-response pairs plays a pivotal role. These pairs provide the model with clear instances of a problem (the prompt) and its desired solution (the response), essentially guiding the model's learning trajectory toward understanding and generating the expected outputs.

In our work, we curated a unique set of prompt-response pairs specifically tailored for data race detection tasks. The "prompt" encapsulates the code segment under consideration, while the "response" comprises the corresponding label information obtained during our label extraction process. This strategy taps into the enormous capacity of LLMs to comprehend intricate text patterns, linking these patterns to labels that signify the presence or absence of a data race.

We systematically produced these pairs using an automated procedure, ensuring extensive coverage of potential data race scenarios and providing a comprehensive learning experience for the LLM. These prompt-response pairs constitute a crucial component of the DataRaceBench-ML dataset, enhancing its effectiveness for fine-tuning LLMs and bolstering its relevance in real-world data race detection tasks.

Incorporating such pairs in our dataset underscores the potential of LLMs to boost the performance of traditional data race detection tools, paving the way for a new era of more efficient, robust, and scalable solutions.


\subsection{DRB-ML Example}
Listing \ref{lst:drb193} shows an example in DRB-ML derived from microbenchmark in Listing \ref{lst:ex193}. We omit the code content for readability. It is worth mentioning that the "line" value in DRB-ML is based on the code without comments.
\begin{lstlisting}[language=JSON, caption=DRB-ML-193.json, label=lst:drb193]
{
  "ID": 193,
  "name": "DRB193-critical-section3-yes.c"
  "DRB_code": "...",
  "trimmed_code": "...",
  "code_len": 425,
  "data_race": 1,
  "data_race_label": "Y3",
  "var_pairs": [pair0]
  "pair0": {
    "name": ["x", "x"],
    "line": [9, 26],
    "col": [7, 7],
    "operation": ["w", "w"]
  }
}

\end{lstlisting}

Listing \ref{prompt} presents a pair of prompts and responses for DRB-ML labels in listing \ref{lst:drb193}. Again, we omit the code content for readability.

\begin{lstlisting}[language=json, caption=prompt-response example for DRB-ML-193, label=prompt]
{
"prompt": "You are an HPC expert. Examine the following code and identify if there's a data race. If a data race is present, specify the variable pairs causing it, along with their line numbers and operations. Code: ...",
"response": "Yes, the provided code exhibits data race issues. The data race is caused by the variable "x" at line 9 and the variable "x" at line 26. Both instances involve write operations."
}
\end{lstlisting}


\section{Conclusion} 
This paper introduced two datasets: the latest version of DataRaceBench and the new DataRaceBench-ML (DRB-ML). While DataRaceBench is designed to evaluate traditional data race detection tools, DRB-ML is a unique dataset engineered to evaluate and enhance Large Language Models (LLMs) applied to data race detection. As the intersection of high-performance computing and machine learning expands, DRB-ML offers a beacon for future research, promising improved precision and real-world applicability in code analysis.

\begin{acks}
Prepared by LLNL under Contract DE-AC52-07NA27344 (LLNL-PROC-853240) and supported by the U.S. Department of Energy, Office of Science, Advanced Scientific Computing Research.
\end{acks}

\bibliographystyle{ACM-Reference-Format}
\bibliography{bugfest23.bib}


\begin{thebibliography}{4}


\ifx \showCODEN    \undefined \def \showCODEN     #1{\unskip}     \fi
\ifx \showDOI      \undefined \def \showDOI       #1{#1}\fi
\ifx \showISBNx    \undefined \def \showISBNx     #1{\unskip}     \fi
\ifx \showISBNxiii \undefined \def \showISBNxiii  #1{\unskip}     \fi
\ifx \showISSN     \undefined \def \showISSN      #1{\unskip}     \fi
\ifx \showLCCN     \undefined \def \showLCCN      #1{\unskip}     \fi
\ifx \shownote     \undefined \def \shownote      #1{#1}          \fi
\ifx \showarticletitle \undefined \def \showarticletitle #1{#1}   \fi
\ifx \showURL      \undefined \def \showURL       {\relax}        \fi
\providecommand\bibfield[2]{#2}
\providecommand\bibinfo[2]{#2}
\providecommand\natexlab[1]{#1}
\providecommand\showeprint[2][]{arXiv:#2}

\bibitem[Chen et~al\mbox{.}(2023)]%
        {drbml}
\bibfield{author}{\bibinfo{person}{Le Chen}, \bibinfo{person}{Whenhao Wu},
  \bibinfo{person}{Stephen Siegel}, \bibinfo{person}{Pei-hung Lin}, {and}
  \bibinfo{person}{Chunhua Liao}.} \bibinfo{year}{2023}\natexlab{}.
\newblock \showarticletitle{DataRaceBench-ML}.
\newblock  (\bibinfo{year}{2023}).
\newblock
\urldef\tempurl%
\url{https://doi.org/10.5072/zenodo.1230296}
\showDOI{\tempurl}


\bibitem[Liao et~al\mbox{.}(2017)]%
        {liao2017dataracebench}
\bibfield{author}{\bibinfo{person}{Chunhua Liao}, \bibinfo{person}{Pei-Hung
  Lin}, \bibinfo{person}{Joshua Asplund}, \bibinfo{person}{Markus Schordan},
  {and} \bibinfo{person}{Ian Karlin}.} \bibinfo{year}{2017}\natexlab{}.
\newblock \showarticletitle{DataRaceBench: a benchmark suite for systematic
  evaluation of data race detection tools}. In
  \bibinfo{booktitle}{\emph{Proceedings of the International Conference for
  High Performance Computing, Networking, Storage and Analysis}}.
  \bibinfo{pages}{1--14}.
\newblock


\bibitem[Lin and Liao(2021)]%
        {lin2021high}
\bibfield{author}{\bibinfo{person}{Pei-Hung Lin} {and} \bibinfo{person}{Chunhua
  Liao}.} \bibinfo{year}{2021}\natexlab{}.
\newblock \showarticletitle{High-precision evaluation of both static and
  dynamic tools using dataracebench}. In \bibinfo{booktitle}{\emph{2021
  IEEE/ACM 5th International Workshop on Software Correctness for HPC
  Applications (Correctness)}}. IEEE, \bibinfo{pages}{1--8}.
\newblock


\bibitem[Wu et~al\mbox{.}(2023)]%
        {wu2023model}
\bibfield{author}{\bibinfo{person}{Wenhao Wu}, \bibinfo{person}{Jan
  H{\"u}ckelheim}, \bibinfo{person}{Paul~D. Hovland}, \bibinfo{person}{Ziqing
  Luo}, {and} \bibinfo{person}{Stephen~F. Siegel}.}
  \bibinfo{year}{2023}\natexlab{}.
\newblock \showarticletitle{Model Checking Race-Freedom When ``Sequential
  Consistency for Data-Race-Free Programs'' is Guaranteed}. In
  \bibinfo{booktitle}{\emph{Computer Aided Verification}},
  \bibfield{editor}{\bibinfo{person}{Constantin Enea} {and}
  \bibinfo{person}{Akash Lal}} (Eds.). \bibinfo{publisher}{Springer Nature
  Switzerland}, \bibinfo{address}{Cham}, \bibinfo{pages}{265--287}.
\newblock
\showISBNx{978-3-031-37703-7}


\end{thebibliography}










\end{document}